\documentclass[aps,pre,twocolumn,superscriptaddress,showpacs,amsmath,amssymb,footinbib,longbibliography]{revtex4-1}
\usepackage[english]{babel}
\usepackage{amssymb,amsbsy,amsmath}
\usepackage{graphicx}
\usepackage{dcolumn}
\usepackage{bm}
\usepackage{subfigure}
\usepackage{color}
\usepackage{textcomp}
\usepackage[colorlinks,bookmarks=false,citecolor=blue,linkcolor=red,urlcolor=blue]{hyperref}

\newcommand{\op}[1]{%
    \fontdimen12\textfont3=2pt\fontdimen12\scriptfont3=1.4pt%
    \!\null\mathop{\vphantom{#1}\smash{#1}}\limits_{\sim}\null\!}
\newcommand{\xref}[1]{\protect\ref{#1}}
\newcommand{\figref}[1]{Fig.~\protect\ref{#1}}
\newcommand{\fmref}[1]{(\protect\ref{#1})}
\def\bra#1{\langle \, {#1} \, | \,}
\def\ket#1{\, | \, {#1} \, \rangle}
\newcommand{\braket}[2]{\langle \, {#1} \, | \, {#2} \, \rangle}

\renewcommand{\eqref}[1]{Eq.~(\protect\ref{#1})}

\begin{document}
\title{Finite-size scaling of typicality-based estimates}

\author{J\"urgen Schnack}
\email{jschnack@uni-bielefeld.de}
\affiliation{Fakult\"at f\"ur Physik, Universit\"at Bielefeld, Postfach 100131, D-33501 Bielefeld, Germany}
\author{Johannes Richter}
\email{Johannes.Richter@physik.uni-magdeburg.de}
\affiliation{Institut f\"ur Theoretische Physik, Universit\"at Magdeburg, P.O. Box 4120, D-39016 Magdeburg, Germany}
\affiliation{Max-Planck-Institut f\"{u}r Physik Komplexer Systeme,
        N\"{o}thnitzer Stra{\ss}e 38, 01187 Dresden, Germany}
\author{Tjark Heitmann}
\author{Jonas Richter}
\author{Robin Steinigeweg}
\email{rsteinig@uos.de}
\affiliation{Fachbereich Physik, Universit\"at
  Osnabr\"uck, Barbarastr. 7, D-49076 Osnabr\"uck, Germany}

\date{\today}

\begin{abstract}
  According to the concept of typicality, an ensemble average can be accurately
  approximated by an expectation value with respect to a single pure state drawn 
  at random from a high-dimensional Hilbert space. This random-vector 
  approximation, or trace estimator, provides a powerful approach to, e.g., 
  thermodynamic quantities for systems with large Hilbert-space
  sizes, which usually cannot be treated  
  exactly, analytically or numerically. Here, we discuss the finite-size scaling 
  of the accuracy of such trace estimators from two perspectives. First, we 
  study the full probability distribution of random-vector expectation 
  values and, second, the full temperature dependence of the standard deviation. 
  With the help of numerical examples, we find pronounced Gaussian probability 
  distributions and the expected decrease of the standard deviation with system 
  size, at least above certain system-specific temperatures. Below and in 
  particular for temperatures smaller than the excitation gap, simple rules are 
  not available.
\end{abstract}

\keywords{Spin systems, Observables, Trace estimators, Typicality}

\maketitle

\section{Introduction}

Methods such as the finite-temperature Lanczos method (FTLM)
\cite{JaP:PRB94,JaP:AP00,ZST:PRB06,ScW:EPJB10,HaS:EPJB14,PRE:COR17,ScT:PR17},
that rest on trace estimators
\cite{Ski:88,Hut:CSSC89,DrS:PRL93,JaP:PRB94,SiR:IJMPC94,GoM:Stanford97,WWA:RMP06,AvT:ACM11,RoA:FCM15,SAI:NM17}
and thus -- in more modern phrases -- on the idea of typicality
\cite{IMN:IEEE19,SuS:PRL12,SuS:PRL13,OAD:PRE18},
approximate equilibrium thermodynamic observables with very high accuracy 
\cite{JaP:AP00,SRS:arXiv19}.
In the canonical ensemble, the observable can be evaluated
either with respect to a single random vector $\ket{r}$,
\begin{eqnarray}
\label{E-2-A}
O^{\text{r}}(T)
&\approx&
\frac{\bra{r}\op{O}e^{-\beta \op{H}}\ket{r}}
     {\bra{r}e^{-\beta \op{H}}\ket{r}}
\ ,
\end{eqnarray}
or with respect to an average over $R$ random vectors,
\begin{eqnarray}
\label{E-2-C}
O^{\text{FTLM}}(T)
&\approx&
\frac{\sum_{r=1}^R\;\bra{r}\op{O}e^{-\beta \op{H}}\ket{r}}
     {\sum_{r=1}^R\;\bra{r}e^{-\beta \op{H}}\ket{r}}
\ ,
\end{eqnarray}
where numerator and denominator are averaged with respect to the
same set of random vectors. The components of
$\ket{r}$ with respect to an orthonormal basis are taken from a
Gaussian distribution with zero mean (Haar measure
\cite{CoS:CMP06,BaG:PRL09,Rei:NC16}), but in practice other
distributions work as well. $T$, $\beta$, and $\op{H}$
denote the temperature, inverse temperature and
the Hamiltonian, respectively.

In this work, we discuss the accuracy of Eqs. (\ref{E-2-A}) and (\ref{E-2-C}), 
where we particularly focus on the dependence of this accuracy on the system 
size or, to be more precise, the dimension of the effective Hilbert space 
spanned by thermally occupied energy eigenstates. While it is well established 
that the accuracy of both equations increases with the square root of this 
dimension, we shed light on the 
size dependence from two less studied perspectives. First, we study the full 
probability distribution of random-vector expectation values, for the 
specific example of magnetic susceptibility and heat capacity in quantum spin 
systems on a one-dimensional lattice. At high temperatures, our 
numerical simulations unveil that these distributions are remarkably well 
described by simple Gaussian functions over several orders of magnitudes. 
Moreover, they clearly narrow with the inverse square root of the 
Hilbert-space dimension towards a $\delta$ function. Decreasing temperature at 
fixed system size, we find the development of broader and 
asymmetric distributions. Increasing the system size at fixed
temperature, however, distributions become narrow and symmetric
again.
Thus, the mere knowledge of the standard deviation turns out to
be sufficient to describe the full statistics of random-vector
expectation values -- at least at not too low temperatures.

The second central perspective of our work is taken by performing a systematic 
analysis of the scaling of the standard deviation with the system size, over 
the entire range of temperature and in various quantum spin models including 
spin-$1/2$ and spin-$1$ Heisenberg chains, critical spin-1/2 sawtooth chains, 
as well as cuboctahedra with spins $3/2$, $2$, and $5/2$. We show a monotonous 
decrease of the standard deviation with increasing effective Hilbert-space 
dimension, as long as temperature is high compared to some system-specific 
low-energy scale. Below this scale, the scaling can become unsystematic if 
only a very few low-lying energy eigenstates contribute. However, 
when averaging according to Eq.\ (\ref{E-2-C}) over a decent 
number ($\sim 100$) of random vectors, one can still determine the 
thermodynamic average very accurately in all examples considered by us. A quite
interesting example constitutes the critical spin-1/2 sawtooth chain, where 
a single state drawn at random is enough to obtain this average down to very 
low temperatures.

This paper is organized as follows. In Sec.\ \ref{sec-2} we briefly
recapitulate models, methods, as well as typicality-based estimators. In 
Sec.\ \ref{sec-3} we present our numerical examples both for frustrated and 
unfrustrated spin systems. The paper finally closes with a summary and 
discussion in Sec.\ \ref{sec-4}.

\section{Method}
\label{sec-2}

In this article we study several spin systems at zero magnetic
field. They are of finite size and described by the Heisenberg
model, 
\begin{eqnarray}
\label{E-2-1}
\op{H}
&=&
\sum_{i<j}\;
J_{ij}
\op{\vec{s}}_i \cdot \op{\vec{s}}_j
\ ,
\end{eqnarray}
where the sum runs over ordered pairs of spins. Here and in the
following operators are marked by a tilde, i.e. $\op{\vec{s}}_i$
denotes the spin-vector operator at site $i$. $J_{ij}$ denotes the
exchange interaction between a spin at site $i$ and a spin at
site $j$. With the sign convention in \fmref{E-2-1}, $J_{ij}>0$
corresponds to antiferromagnetic interaction.

Numerator and denominator of \fmref{E-2-C}, the latter is the
partition function, are evaluated using a Krylov-space
expansion, i.e.\ a spectral representation of the exponential in
a Krylov space with $\ket{r}$ as starting vector of the 
Krylov-space generation, compare \cite{JaP:PRB94,ScW:EPJB10}.
One could equally well employ Chebyshev polynomials 
\cite{TalEzer1984,Dobrovitski2003,WWA:RMP06} or integrate  the
imaginary-time Schr\"odinger 
equation with a Runge-Kutta method
\cite{SGB:PRL14,SGB:PRB15,ElF:PRL13},
the latter is used later in this paper as well. 

If the Hamiltonian $\op{H}$ possesses symmetries, they can be
used to block-structure the Hamiltonian matrix according to the
irreducible representations of the employed symmetry groups
\cite{ScW:EPJB10,HaS:EPJB14}, which yields for the partition
function 
\begin{eqnarray}
\label{E-2-D}
Z^{\text{FTLM}}(T)
&\approx&
\sum_{\gamma=1}^\Gamma\;
\frac{\text{dim}[{\mathcal H}(\gamma)]}{R}
\nonumber
\\
&&
\times
\sum_{r=1}^R\;
\sum_{n=1}^{N_L}\;
e^{-\beta \epsilon_n^{(r)}} |\braket{n(r)}{r}|^2
\ .
\end{eqnarray}
${\mathcal H}(\gamma)$ labels the subspace that belongs to 
the irreducible representation $\gamma$, $N_L$ denotes the dimension
of the Krylov space, and $\ket{n(r)}$ is the n-th
eigenvector of $\op{H}$ in this Krylov space grown from
$\ket{r}$. The energy eigenvalue is $\epsilon_n^{(r)}$.
To perform the Lanczos
diagonalization for larger system sizes, we use
the public 
code {\it spinpack} \cite{spin:256,RiS:EPJB10}.

In our numerical studies we evaluate the uncertainty of a
physical quantity by repeating its numerical evaluation
$N_S$ times. For this statistical sample we define the standard 
deviation of the observable in the following way:
\begin{eqnarray}
\label{E-2-E}
\delta(O)
&=&
\sqrt{
\frac{1}{N_S}
\sum_{r=1}^{N_S}\;
\left[
O^{\text{m}}(T)
\right]^2
-
\left[
\frac{1}{N_S}
\sum_{r=1}^{N_S}\;
O^{\text{m}}(T)
\right]^2
}
\nonumber
\\
&=&
\sqrt{
\overline{\left[
O^{\text{m}}(T)
\right]^2}
-
\left[\overline{
O^{\text{m}}(T)}
\right]^2
}
\ .
\end{eqnarray}
$O^{\text{m}}(T)$ is either evaluated according to
\eqref{E-2-A} (m=r) or to \eqref{E-2-C} (m=FTLM), depending on whether the
fluctuations of approximations with respect to one random vector
or with respect to an average over $R$ vectors shall be
investigated.

We consider two physical quantities, the zero-field
susceptibility as well as the heat capacity. Both are evaluated
as variances of magnetization and energy, respectively, i.e. 
\begin{eqnarray}
\label{E-2-24}
\chi(T)
&=&
\frac{(g \mu_B)^2}{k_B T}\,
\Bigg [
\left\langle(\op{S}^z)^2\right\rangle
-
\left\langle\op{S}^z\right\rangle^2
\Bigg ] \, ,
\\
C(T)
&=&
\frac{k_B}{(k_B T)^2}\,
\Bigg [
\left\langle\op{H}^2\right\rangle
-
\left\langle\op{H}\right\rangle^2
\Bigg ]
\ .
\end{eqnarray}
We compare our results with the well-established
high-temperature estimate 
\begin{eqnarray}
\label{E-1-3}
{\delta \langle\op{O}\rangle}
&\simeq&
{\langle\op{O}\rangle}\frac{\alpha}{\sqrt{Z_{\text{eff}}}}
\ ,\qquad
Z_{\text{eff}} = \text{tr}\left[e^{-\beta (\op{H}-E_0)}\right]
\ .
\end{eqnarray}
Here $E_0$ denotes the ground-state energy.
In general the prefactor $\alpha$ depends on the specific
system, its structure and size, as well as on temperature
\cite{SuS:PRL12,SuS:PRL13}, but empirically often turns out to 
be a constant of order $\alpha\approx 1$ for high enough
temperatures, compare \cite{JaP:AP00,PRE:COR17,SRS:arXiv19}.
Rigorous error bounds, see Refs.~\cite{HaD:PRE00,SuS:PRL13},
share the dependence on $1/\sqrt{Z_{\text{eff}}}$, but lead to a
prefactor that can be substantially larger than the
empirical finding.

\section{Numerical results}
\label{sec-3}

We now present our numerical results. First, in the following
Sec.~\ref{sec-3-0}, the full probability distribution of random-vector
expectation values is discussed for shorter
spin chains, where this distribution can be easily obtained by
generating a large set of different random vectors.
In the remainder of Sec.~\ref{sec-3} the size dependence of the
standard deviation is investigated for longer
spin chains of spin $s=1/2$ and $s=1$, respectively, which are
treated by Lanczos methods. The interesting behavior of a
quantum critical delta chain is studied as well. Finally, we
discuss the dependence of the standard deviation on the spin quantum
number for a body of fixed size, the cuboctahedron.

\subsection{Distribution of random-vector expectation values for smaller 
antiferromagnetic spin-$1/2$ chains}\label{sec-3-0}

As a first step, in order to judge the accuracy of the 
single-state estimate in \eqref{E-2-A}, it is instructive to study its 
full probability distribution $p$, obtained by drawing many [here ${\cal 
O}(10^4 - 10^6)$] random vectors. To be more precise, we
evaluate the numerator of \eqref{E-2-A}  
for different random states $\ket{r}$, while its denominator is
calculated as  
the average over all $\ket{r}$,
\begin{equation}
\frac{\bra{r}\op{O}e^{-\beta \op{H}}\ket{r}}
     {\sum_{r=1}^R\;\bra{r}e^{-\beta \op{H}}\ket{r}}
\ . \label{eq_new}
\end{equation}
The advantage of using this equation, instead of \eqref{E-2-A},
is that the mean coincides with \eqref{E-2-C}, the latter should be used
to correctly obtain the low-temperature average in system of finite size
\cite{SRS:arXiv19}. However, at sufficiently high temperatures or in
sufficiently large systems, one might equally well use \eqref{E-2-A}, as we
have checked.

The single results for \eqref{eq_new} are then collected into 
bins of appropriate width in order to form a ``smooth'' distribution $p$. While 
one might expect that $p$ will be approximately symmetric around the 
respective thermodynamic average, the width of the distribution 
indicates how reliable a single random vector can approximate the ensemble 
average.

\begin{figure}[tb]
\centering
\includegraphics*[clip,width=1\columnwidth]{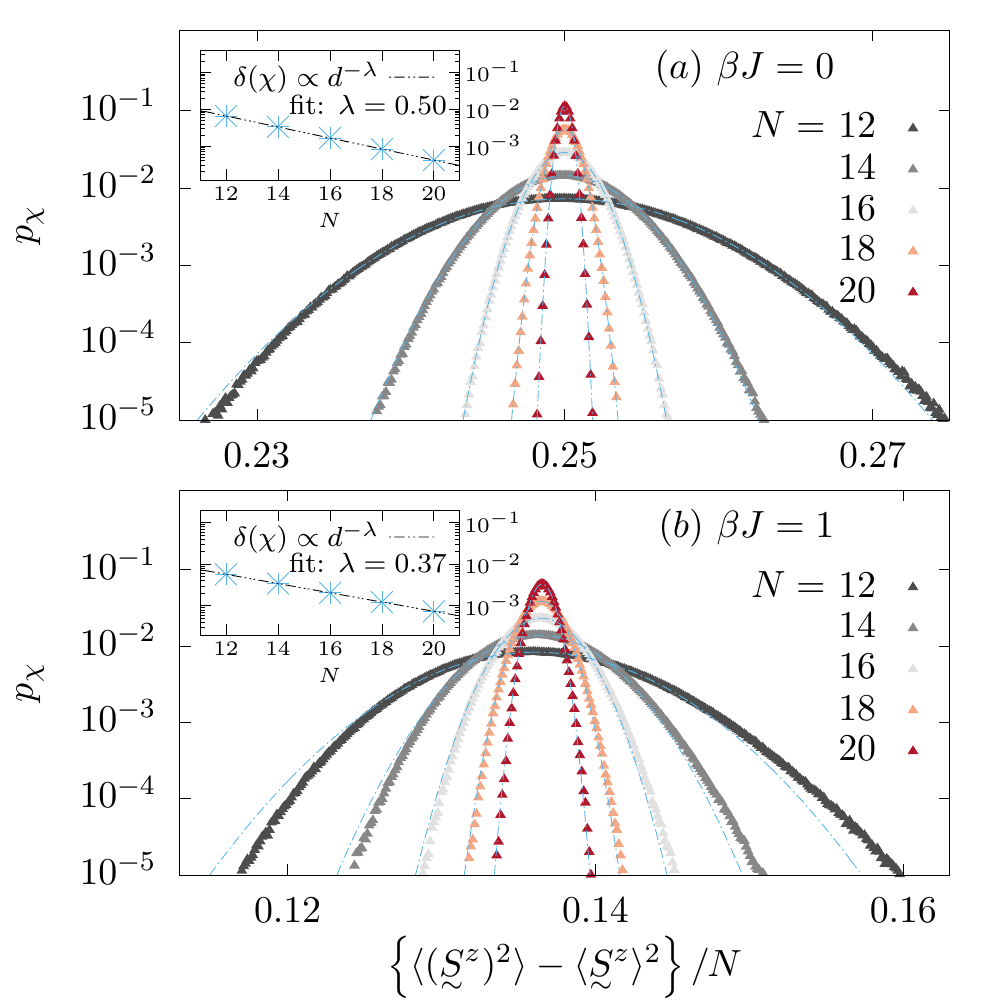}
\caption{(a) Probability distribution of the susceptibility 
$\chi(T)T/N$ evaluated from independently drawn single states 
according to \eqref{eq_new}. Data is shown for different system sizes $N = 
12,\dots,20$ at infinite temperature $\beta J = 0$. The dashed lines indicate 
Gaussian fits to the data. The inset shows the standard deviation 
$\delta(\chi)$ versus $N$, which scales as $\delta(\chi)\propto1/\sqrt{d}$ 
with Hilbert-space dimension $d=2^N$. (b) Same data as in (a) but now for the finite 
temperature $\beta J = 1$.}
\label{ftlm-scaling-A}
\end{figure}

\begin{figure}[tb]
\centering
\includegraphics*[clip,width=1\columnwidth]{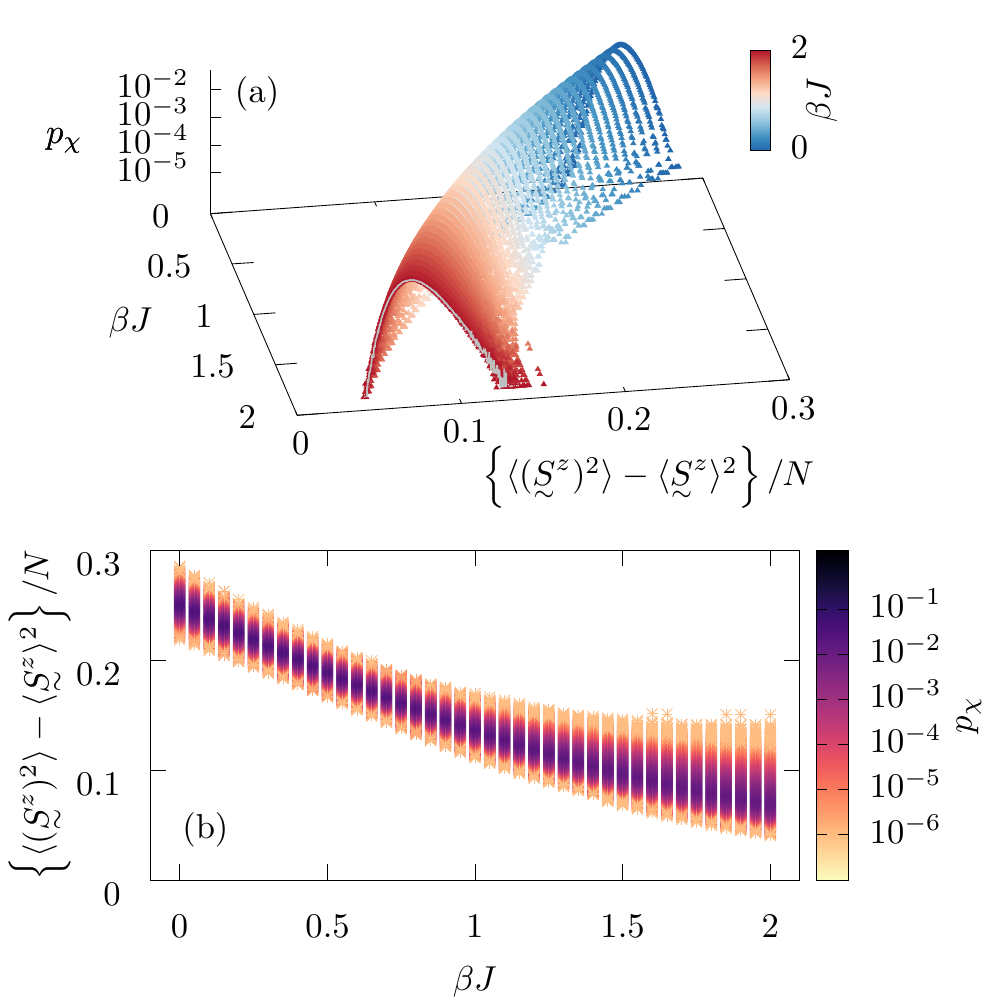}
\caption{(a) Probability distribution of the susceptibility 
$\chi(T)T/N$ for various temperatures $0 \leq \beta J \leq 2$ at the fixed 
system size $N = 12$ obtained by ED (symbols). For comparison,
data obtained by Runge-Kutta at $\beta J=2$ is shown as well
(curve). (b) Same data as in (a), but now as a contour plot.} 
\label{ftlm-scaling-B}
\end{figure}

In this Section, we study the probability distribution $p$ (in the following 
denoted as $p_\chi$ and $p_C$) for the 
quantities $\chi(T)T/N$ and $C(T)T^2/N$, and exemplarily
consider the one-dimensional  
spin-$1/2$ Heisenberg model with antiferromagnetic nearest-neighbor coupling 
$J > 0$ and chain length $N$. Note that, as discussed in the
upcoming Secs.~\ref{sec-3-1} - \ref{sec-3-4}, details of the
model can indeed have an impact  
on the behavior of $p$ in certain temperature regimes. Note further, that we 
focus in this Section on small to intermediate system sizes $N \leq 20$, 
where $p$ can be easily obtained by generating a large set of
different random vectors and evolving these vectors in imaginary time by, e.g., a
simple Runge-Kutta scheme. We have checked that the Runge-Kutta scheme employed 
in this Section has practically no impact on $p$.

To begin with, in \figref{ftlm-scaling-A}~(a), $p_\chi$ is shown for 
different chain lengths $N = 12,\dots,20$ at infinite
temperature $\beta J = 0$.
For all values of $N$ shown here, we find that $p_\chi$ is well described 
by a Gaussian distribution \cite{Reimann2019} over several orders of magnitude. 
While the mean of these Gaussians is found to accurately coincide with the 
thermodynamic average $\lim_{T \to \infty}\chi(T)T/N = 
1/4$ \cite{SSL:PRB01}, we moreover observe that the width of the Gaussians becomes  
significantly narrower upon increasing $N$. This fact already visualizes that 
the accuracy of the estimate in \eqref{E-2-A} improves for increasing 
Hilbert-space dimension. In particular, as shown in the inset of 
\figref{ftlm-scaling-A}, the standard deviation $\delta(\chi)$ scales as 
$\delta(\chi)\propto 
1/\sqrt{d}$, where $d = 2^N$ is the dimension of the Hilbert space. This is in 
agreement with \eqref{E-1-3} for $\alpha\approx1.2$ and $Z_\text{eff} = d$ at $\beta = 0$. 
Note that since $p_\chi$ is found to be a Gaussian, the width
$\delta(\chi)$ is sufficient to describe the whole distribution (apart from the 
average).

To proceed, \figref{ftlm-scaling-A}~(b) again shows the probability 
distribution $p_\chi$, but now for the finite temperature $\beta J = 1$. 
There are two important observations compared to the previous case of $\beta J 
= 0$. 
First, for small $N = 12$, one clearly finds that $p_\chi$ now takes on an 
asymmetric shape and the tails are not described by a Gaussian
anymore. Importantly, however, upon increasing the system size
$N$, $p_\chi$ becomes narrower and eventually turns into a
Gaussian again. One may speculate about possible reasons for the
observed asymmetry: It might reflect an asymmetry of the
distribution, which is already present at $\beta = 0$ and small
$N$, and then increases with increasing $\beta$; or
it might also stem from the boundedness (positivity) of the
observables, although 
the bounds are still far away for the presented case of $\beta J
= 1$ in \figref{ftlm-scaling-A}~(b). While this asymmetry remains
to be explored in more detail in future work, it is expected
that the Gaussian shape breaks down in small dimensions of the
effective Hilbert space dimensions \cite{Reimann2019}. It is
worth pointing out that, even 
for very large dimensions, the very outer tails of the distribution are
expected to be of non-Gaussian nature \cite{Reimann2019}. Yet, these
tails are hard to resolve in our numerical simulations, since a huge number
of samples would be needed.

As a second difference compared to $\beta J = 0$, we find that 
although 
$p_\chi$ becomes narrower for larger $N$ also at $\beta J = 1$, this scaling 
is now considerably slower as a function of dimension $d$ (see inset of 
\figref{ftlm-scaling-A}~(b)). This is caused by the smaller effective 
Hilbert-space dimension $Z_\text{eff} < d$ at $\beta J > 0$. 
As a consequence, for a fixed value of $N$, the 
single-state estimate in \eqref{E-2-A} becomes less reliable at $\beta J = 
1$ compared to $\beta J= 0$. However, let us stress that accurate 
calculations are still 
possible at $T > 0$ as long as $N$ is sufficiently large.
(Recall, that $N \leq 20$ was chosen to be able to generate a
large set of random vectors.)

\begin{figure}[tb]
\centering
\includegraphics*[clip,width=1\columnwidth]{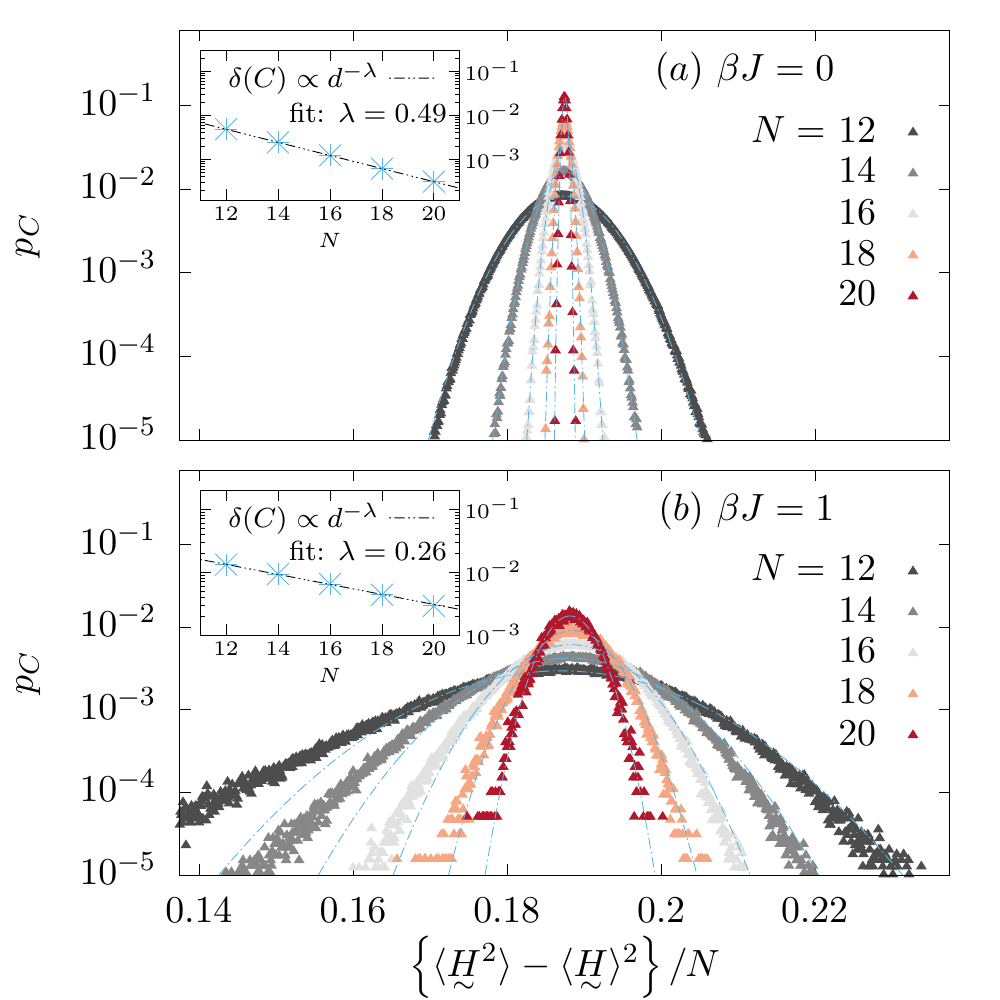}
\caption{Analogous data as in \figref{ftlm-scaling-A}, but now 
for the heat capacity $C(T)T^2/N$.}
\label{ftlm-scaling-C}
\end{figure}

\begin{figure}[tb]
\centering
\includegraphics*[clip,width=1\columnwidth]{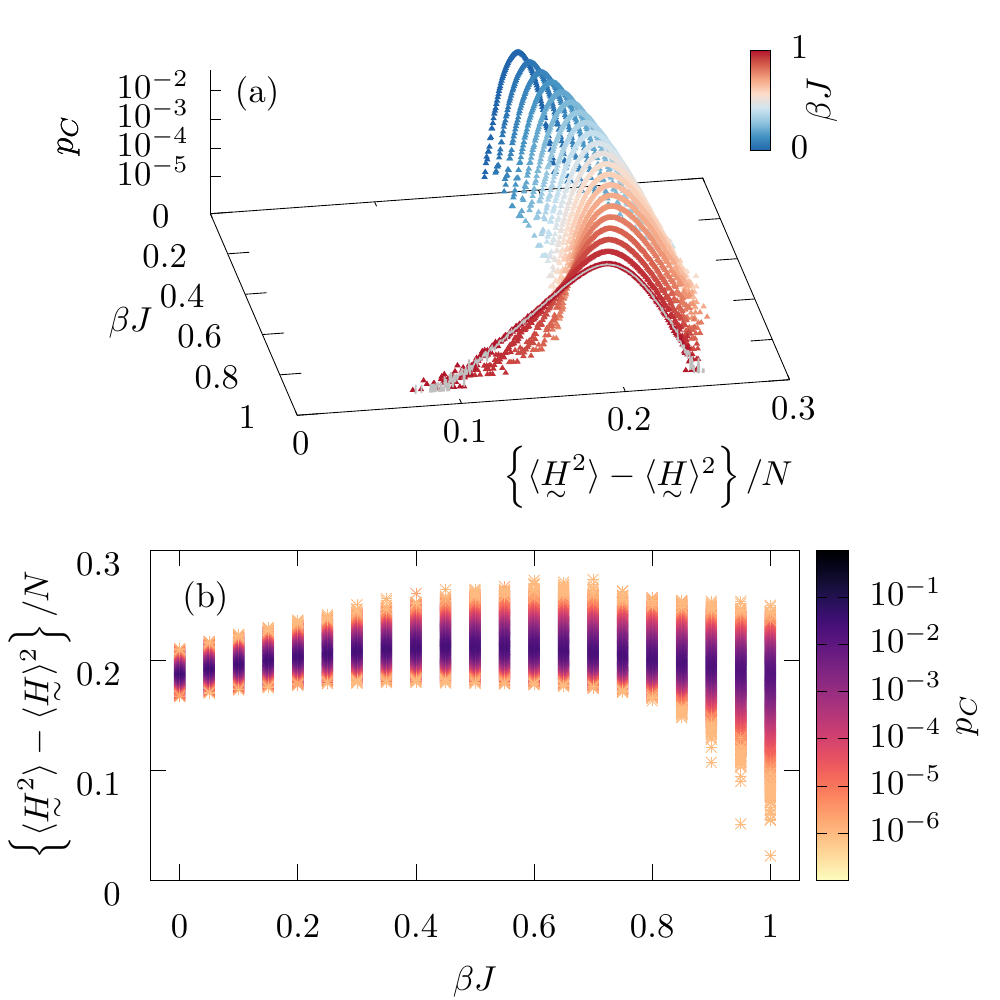}
\caption{Analogous data as in \figref{ftlm-scaling-B}, but now 
for the heat capacity $C(T)T^2/N$.}
\label{ftlm-scaling-D}
\end{figure}

In order to analyze the development of the probability 
distribution with respect to temperature in more detail, 
\figref{ftlm-scaling-B}~(a) shows $p_\chi$ for various values of $\beta J$ in 
the range $0 \leq \beta J \leq 2$, for a fixed small system size
$N = 12$. Note 
that the qualitative behavior in principle is independent of $N$, but better to
visualize for small $N$ with a broader $p_{\chi}$. Starting from the 
high-temperature limit $\lim_{T \to \infty}\chi(T)T/N = 1/4$, we
find that the maximum of $p_\chi$ gradually shifts towards
smaller  values upon decreasing $T$.

This shift of the maximum is clearly 
visualized also in \figref{ftlm-scaling-B}~(b), which shows the same data, but in 
a different style. Moreover, 
\figref{ftlm-scaling-B}~(b) additionally highlights the fact that the 
probability distribution $p_\chi$ for a fixed (and small) value of $N$ becomes 
broader (and asymmetric) for intermediate values of $T$. Note, that $p_\chi$ 
might become narrower again for smaller values of $T$, see also discussion in 
Secs.\ \ref{sec-3-1} - \ref{sec-3-2}.

Eventually, in \figref{ftlm-scaling-C} and 
\figref{ftlm-scaling-D}, we present analogous results for the full probability 
distribution $p$, but now for the heat capacity $C(T)T^2/N$. Overall, our 
findings for $p_C$ are very similar compared to the previous discussion of 
$p_\chi$. Namely, we find that at $\beta J = 0$, $p_C$ is very well described 
by Gaussians over several orders of magnitude. Moreover, the standard deviation 
$\delta(C)$ again scales as $\propto 1/\sqrt{d}$ at $\beta = 0$. As shown 
in \figref{ftlm-scaling-C}~(b) and also in \figref{ftlm-scaling-D}, the 
emerging asymmetry of the probability distribution at small $N$ and finite $T$ 
is found to be even more pronounced for the heat capacity compared to the 
previous results for $\chi(T)$. Interestingly, we find that the maximum of 
$p_C$, on the contrary, displays only a minor dependence on temperature (at 
least for the values of $\beta J$ shown in \figref{ftlm-scaling-D} - naturally, 
it is expected to change at lower temperatures and will to go to zero at 
temperature $T=0$).

\subsection{Larger antiferromagnetic spin-$1/2$ chains}
\label{sec-3-1}

Using a Krylov-space expansion one can nowadays reach large
system sizes of $N\in [40,50]$ for spins $s=1/2$, see
e.g.\ \cite{SSR:PRB18}. But since we also perform a statistical
analysis we restrict calculations to $N\leq 36$ spins.

Following the scaling behavior of
$\{\langle(\op{S}^z)^2\rangle
-
\langle\op{S}^z\rangle^2\}$
as well as
$
\{\langle\op{H}^2\rangle
-
\langle\op{H}\rangle^2\}$,
which is shown in
Figs.~\xref{ftlm-scaling-A} and \xref{ftlm-scaling-C},
one expects a very narrow distribution of both quantities for
$N=36$ compared to e.g. $N=20$ since the dimension is
$2^{16}=65536$ times bigger for $N=36$ which yields a 256
times narrower distribution. Such a distribution is smaller than
the linewidth in a plot.

\begin{figure}[ht!]
\centering
\includegraphics*[clip,width=0.69\columnwidth]{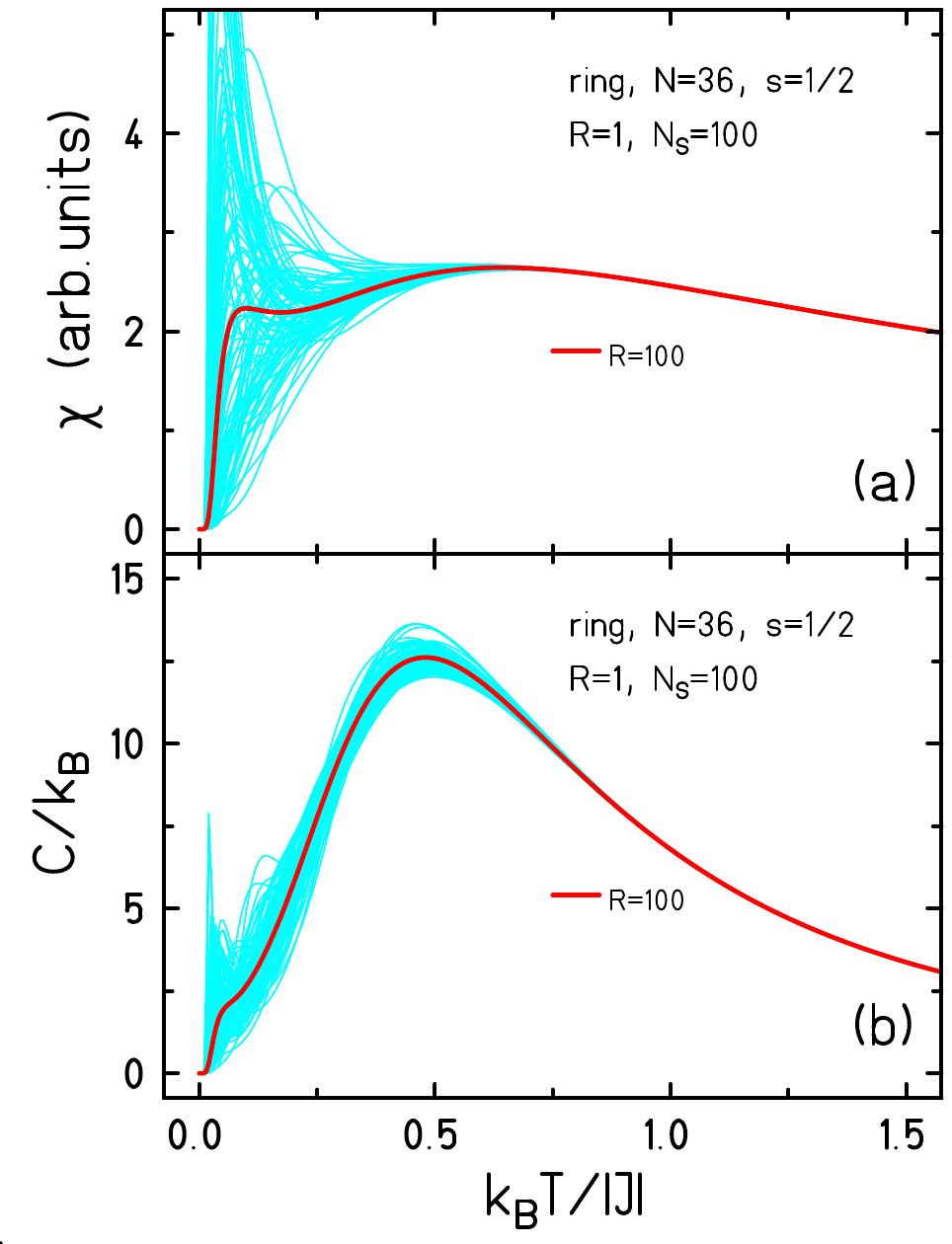}
\caption{Spin ring $N=36$, $s=1/2$: The
  light-blue curves depict 100 different estimates of the
  susceptibility (a) as well as of the heat capacity (b).
  The FTLM estimate for $R=100$ is also presented.}
\label{ftlm-scaling-Z}
\end{figure}

\begin{figure}[ht!]
\centering
\includegraphics*[clip,width=0.69\columnwidth]{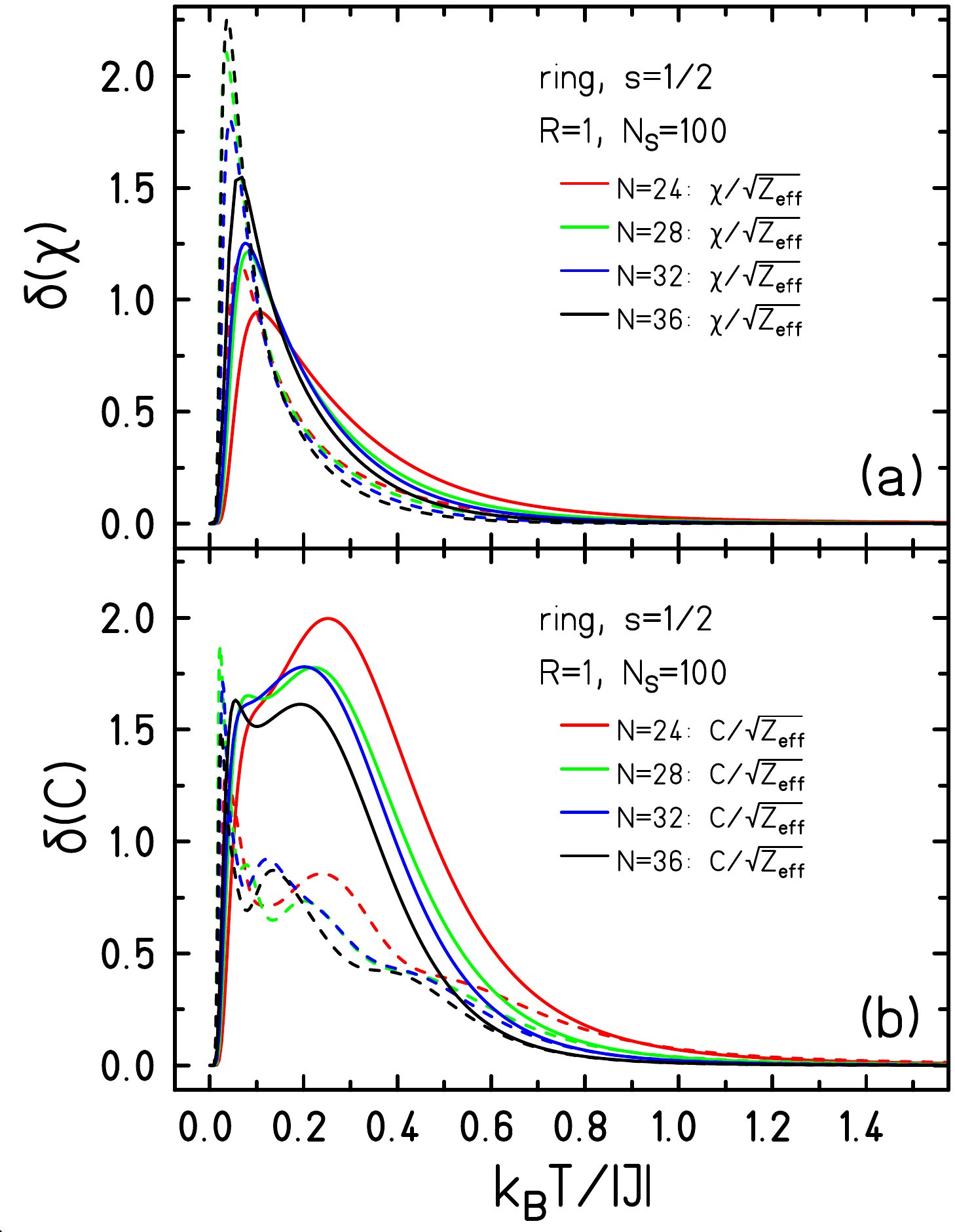}
\caption{Spin rings, $s=1/2$: Computed standard deviations
  (dashed curves) of the susceptibility (a) and the heat
  capacity (b) compared to the error estimate (solid curves) for
  various sizes $N$. The same color denotes the same system.}  
\label{ftlm-accuracy-f-Y}
\end{figure}

That the distributions are narrow can be clearly seen by eye
inspection in \figref{ftlm-scaling-Z} where the light blue
curves depict thermal expectation values according to
\eqref{E-2-A}. For $k_B T > |J|$ they fall on top of each other
and coincide with the average
over $R=100$ realizations. Below this temperature the
distributions widen, which is magnified by the fact that the
real physical quantities susceptibility and heat capacity
contain factors of $1/T$ and $1/T^2$, respectively.

Their standard deviation is provided in
\figref{ftlm-accuracy-f-Y}. Coming from high temperatures, the
universal behavior \fmref{E-1-3} switches to a behavior that
in general depends on system (here chain) and size below a
characteristic temperature, here $k_B T \approx
|J|$. Nevertheless, the qualitative expectation that the
standard deviation shrinks with increasing system size is met
down to $k_B T \approx 0.2 |J|$, below which no definite
statement about the dependence on system size can be made. We
conjecture that with growing $N$ the increasing density of
low-lying states as well as the vanishing excitation gap between
singlet ground state and triplet first excited state influence
the behavior at very low temperatures strongly.

\subsection{Antiferromagnetic spin-$1$ chains}
\label{sec-3-2}

\begin{figure}[ht!]
\centering
\includegraphics*[clip,width=0.69\columnwidth]{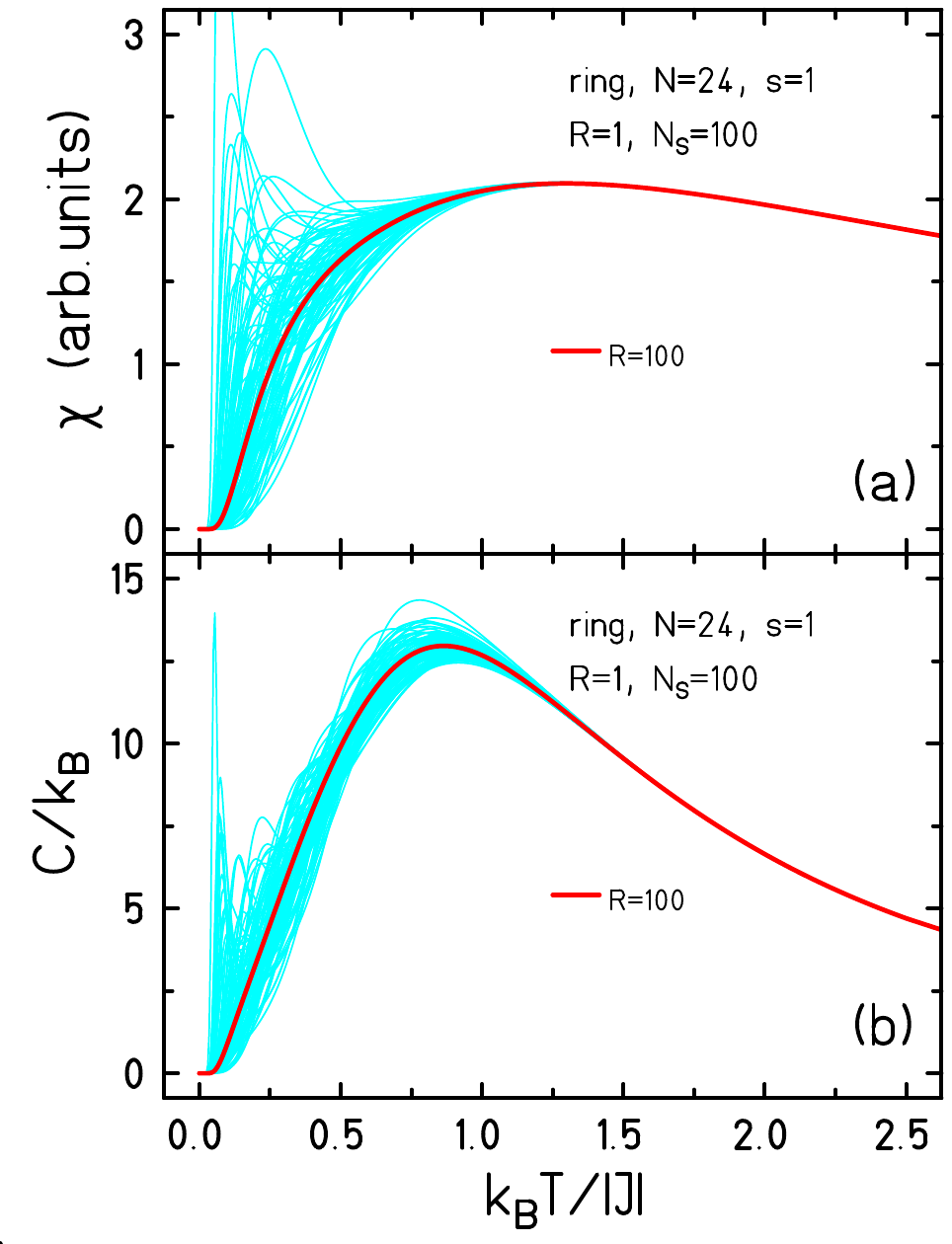}
\caption{Spin ring $N=24$, $s=1$: The
  light-blue curves depict 100 different estimates of the
  susceptibility (a) as well as the heat capacity
  (b).
  The FTLM estimate for $R=100$ is also presented.}
\label{ftlm-accuracy-f-X}
\end{figure}

\begin{figure}[ht!]
\centering
\includegraphics*[clip,width=0.69\columnwidth]{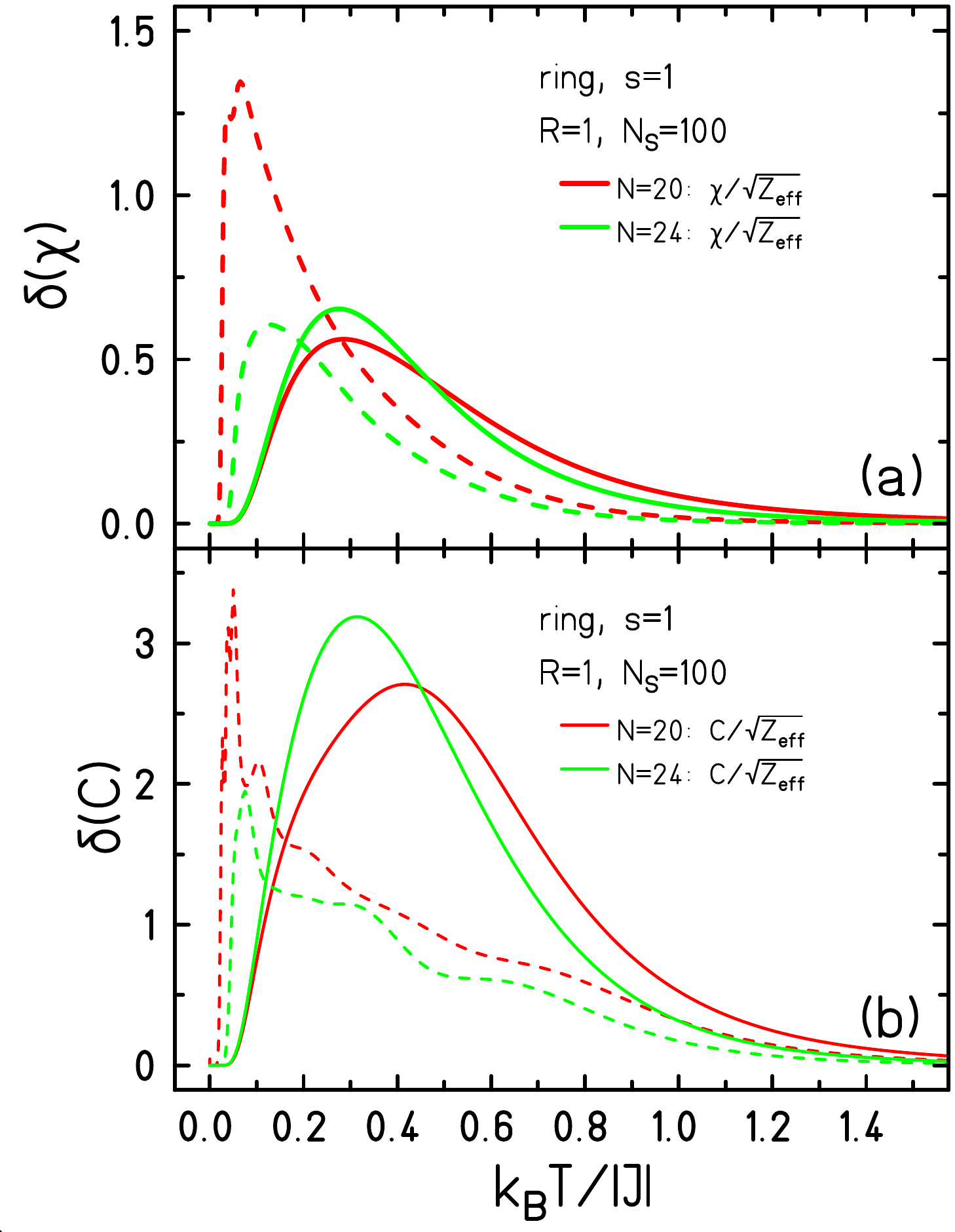}
\caption{Spin rings, $s=1$: Computed standard deviations
  (dashed curves) of the susceptibility (a) and the heat
  capacity (b) compared to the error estimate (solid curves) for
  various sizes $N$. The same color denotes the same system.}
\label{ftlm-accuracy-f-W}
\end{figure}

In order to monitor an example where a vanishing excitation gap
cannot be expected, not even in the thermodynamic limit, we
study spin-1 chains that show a Haldane gap
\cite{Hal:PL83,Hal:PRL83}, see \figref{ftlm-accuracy-f-X}.
The scaling formula \fmref{E-1-3} 
indeed suggests that for $k_B T \lessapprox (0.4\dots 0.5) |J|$ the standard
deviations of the larger system with $N=24$ should exceed those
of the smaller system with $N=20$, 
compare crossing curves of the estimator in
\figref{ftlm-accuracy-f-W}.
However, the actual simulations show that this is not the
case. The low-temperature fluctuations in the gap region are
smaller for the larger system, at least for the two investigated
system sizes.

\subsection{Critical Spin-$1/2$ delta chains}
\label{sec-3-3}

As the final one-dimensional example we investigate a delta
chain (also called sawtooth chain) in the quantum critical
region, i.e., thermally excited above the quantum critical point
(QCP) \cite{KDN:PRB14,DmK:PRB15,BML:npjQM18}. The QCP is met
when the ferromagnetic nearest-neighbor interaction $J_1$ and
the antiferromagnetic next-nearest neighbor interaction $J_2$
between spins on adjacent odd sites assume a ratio of
$|J_2/J_1|=1/2$. At the QCP the system features a
massive ground-state degeneracy due to multi-magnon flat bands
as well as a double-peak density of states
\cite{KDN:PRB14,DmK:PRB15,SRS:arXiv19}. Moreover, 
the  typical
finite-size gap is virtually not present at the QCP \cite{KDN:PRB14}.

\begin{figure}[ht!]
\centering
\includegraphics*[clip,width=0.69\columnwidth]{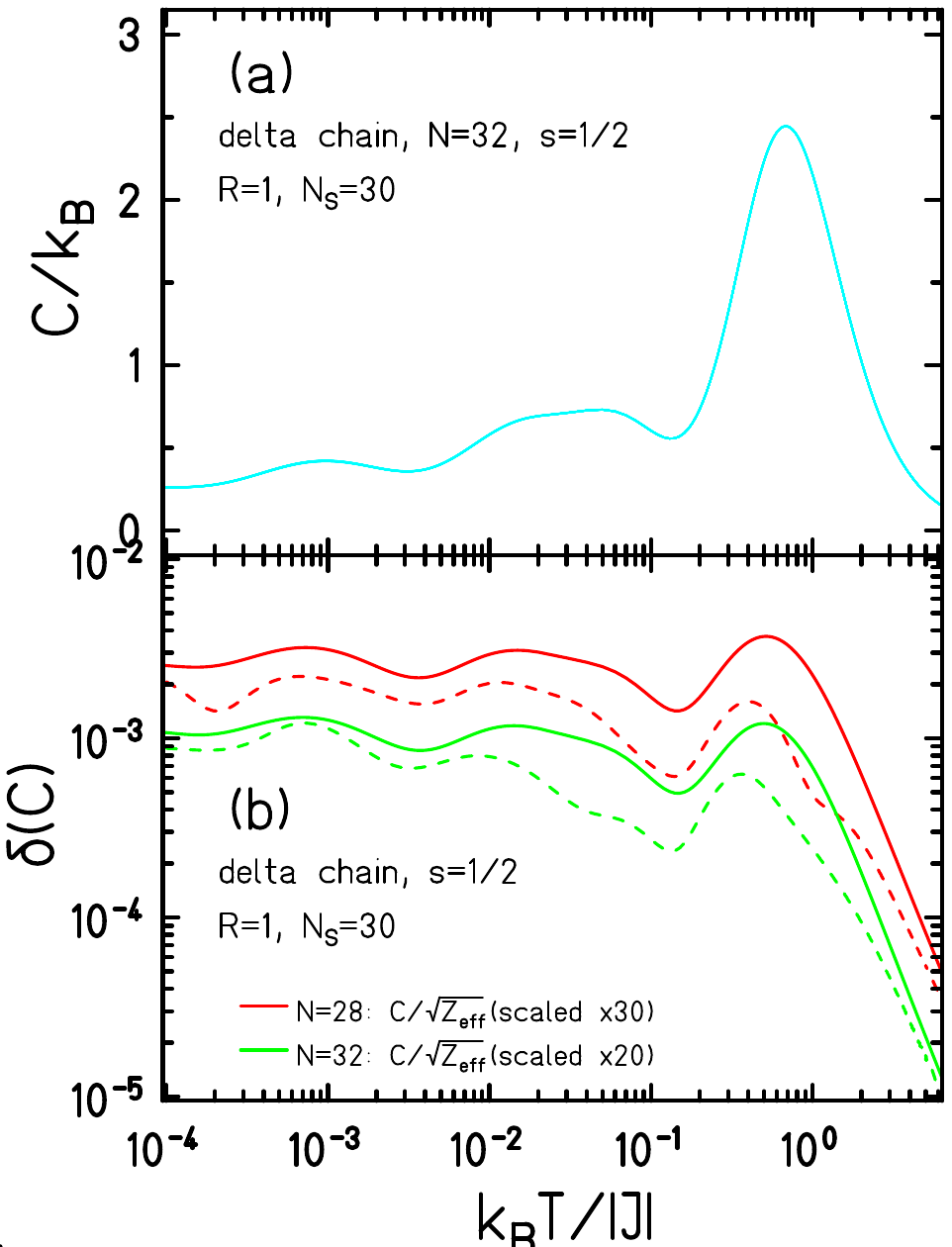}
\caption{Delta chain $s=1/2$, $|J_2/J_1|=0.5$: heat
  capacity for $N=32$ (a) and standard deviation for $N=28$ and
  $N=32$ (b). The light-blue curves depict 
  $N_S=30$ different estimates of the heat capacity (there are indeed 30
  curves in this plot, which are indistinguishable by
  eye). Computed standard deviations (dashed curves) are
  compared to the error estimate (solid curves). The same color
  denotes the same system. 
}
\label{ftlm-accuracy-f-T}
\end{figure}

Since the QCP does not depend on the size of the system and the
structure of the analytically known multi-magnon flat band
energy eigenstates does not either, we do not expect to find
large finite-size effects when investigating the standard
deviation of observables, e.g.\ of the
heat capacity. It turns even out that by eye inspection no
fluctuations are visible in \figref{ftlm-accuracy-f-T}~(a).
The figure shows $N_S=30$ thermal expectation values
\fmref{E-2-1} that virtually fall on top of each other. This
means that a single random vector provides the equilibrium
thermodynamic functions for virtually all temperatures.
When evaluating the standard deviation, dashed curves in
\figref{ftlm-accuracy-f-T}~(b), it turns out that it is
unusually small, even for very low temperatures. The estimator 
\fmref{E-1-3} to which we compare had to be scaled in this case
which might have two reasons. One reason could be that the large
ground state degeneracy cannot be fully captured 
by the Krylov space expansion and thus the evaluation of the
estimator \fmref{E-1-3} by means of \eqref{E-2-D} is
inaccurate. The other reason could be that the empirical finding
of $\alpha\approx 1$ is not appropriate in this special case of
a quantum critical system. However, the general rule that
trace estimators are more accurate in larger Hilbert spaces is
also observed here. The standard deviation of the smaller delta
chain with $N=28$ is a few times larger than for $N=32$. 

The result is an impressive example for what it means that a
quantum critical system does not possess any intrinsic scale
in the quantum critical region \cite{Voj:AdP00,Voj:RPP03}. The
only available scale is temperature. This means in particular
that the low-energy spectrum is dense and
therefore does not lead to any visible fluctuations of the
estimated observables.

\subsection{Antiferromagnetic cuboctahedra with spins $3/2$, $2$, and $5/2$}
\label{sec-3-4}

\begin{figure}[ht!]
\centering
\includegraphics*[clip,width=0.69\columnwidth]{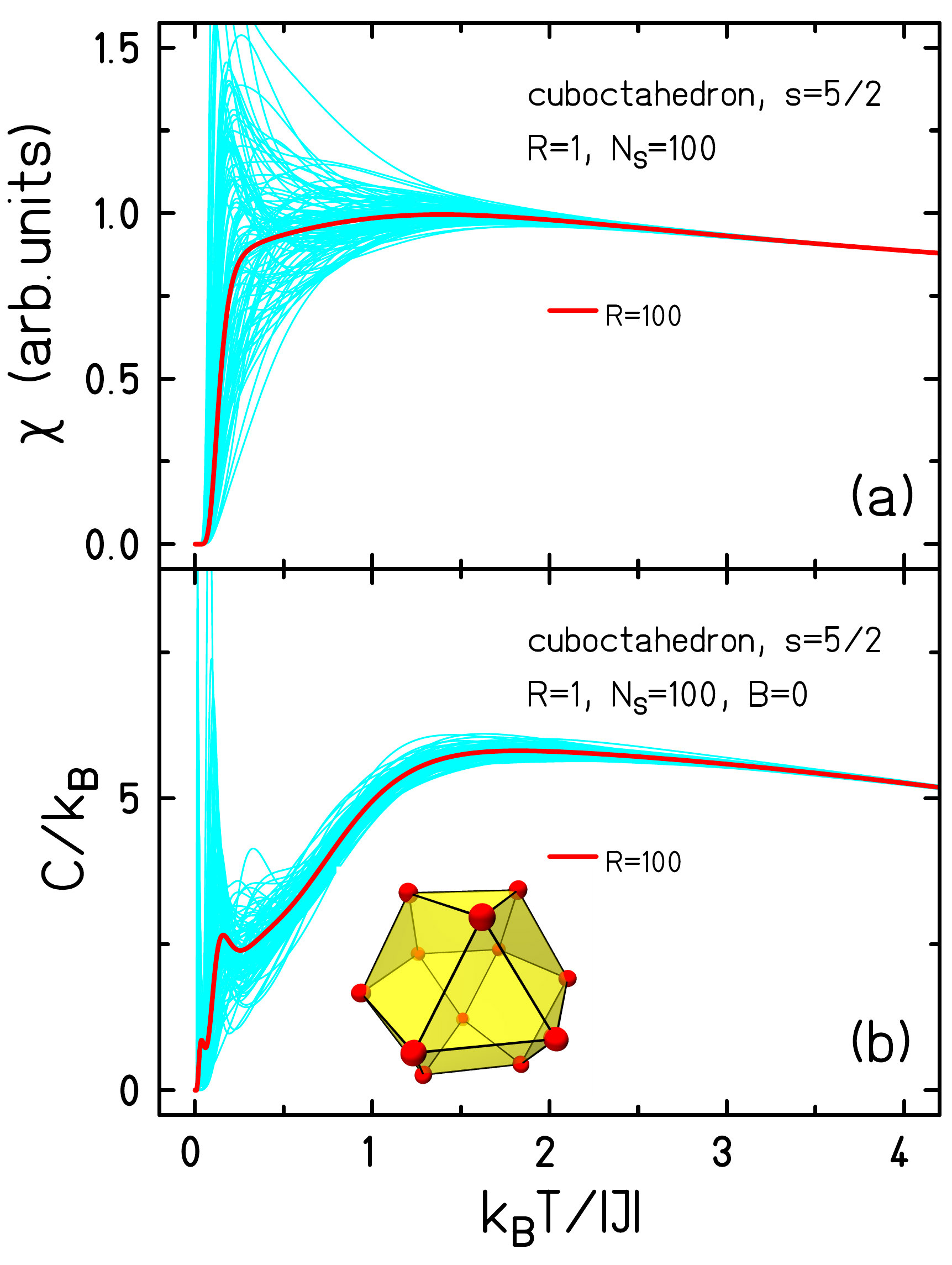}
\caption{Cuboctahedron $N=12$, $s=5/2$: The
  light-blue curves depict 100 different estimates of the
  susceptibility (a) as well as the heat capacity
  (b).
  The FTLM estimate for $R=100$ is also presented. The structure
of the cuboctahedron is displayed in (b).}
\label{ftlm-accuracy-f-V}
\end{figure}

Our last scaling analysis differs from the previous
examples. The cuboctahedron is a finite-size body, that is
equivalent to a kagome lattice with $N=12$
\cite{RLM:PRB08,HoZ:JPCS09,SHL:PRE17}. The structure is shown in
\figref{ftlm-accuracy-f-V}(b). Here, we 
vary the spin quantum number, not the size of the
system. The dimension 
of the respective Hilbert spaces grows considerably which leads
to the expected scaling \fmref{E-1-3} above temperatures of
$k_B T \approx 1.5 |J|$. But the low-temperature behavior, in
particular of the heat capacity for temperatures below the
crossing of the estimators, eludes the expected order of
more accurate results, i.e. smaller fluctuations for larger
Hilbert-space dimension.

While the low-temperature behavior and
the standard deviation of the susceptibility are largely
governed by the energy gap between singlet ground state and
triplet excited state, and this does not vary massively with the
spin quantum number, the heat capacity is subject to stronger
changes. When going from smaller to larger spin quantum numbers
the strongly frustrated spin system looses some of its
characteristic quantum properties while becoming more classical
with increasing spin $s$. In particular, the low-lying singlet states
below the first triplet state which dominate the low-temperature heat
capacity move out of the gap for larger spin $s$
\cite{SSR:JMMM05,ScS:P09}. 

\begin{figure}[ht!]
\centering
\includegraphics*[clip,width=0.69\columnwidth]{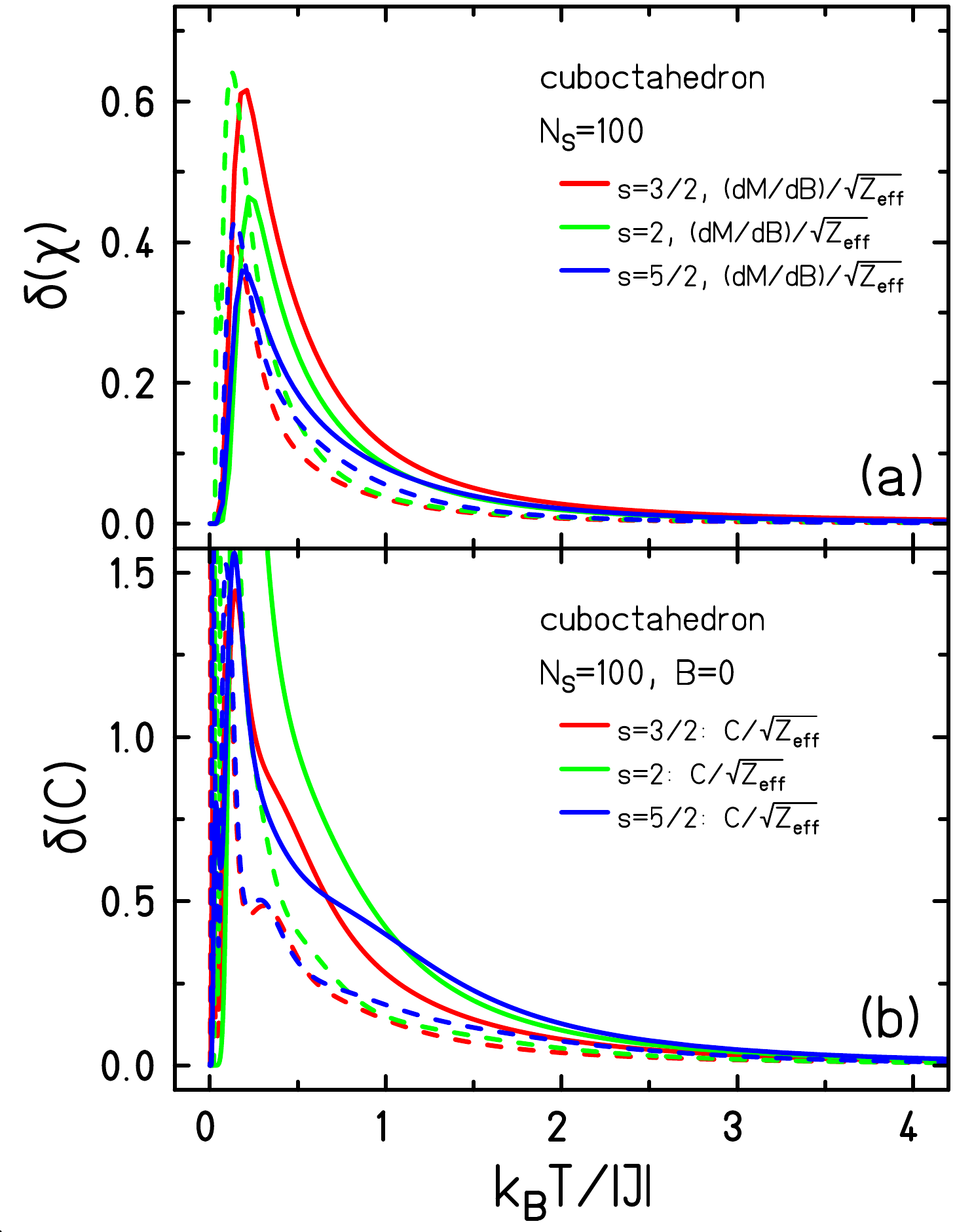}
\caption{Cuboctahedron $N=12$: Computed standard deviations
  (dashed curves) of the susceptibility (a) and the heat
  capacity (b) compared to the error estimate (solid curves) for
  various spin quantum numbers $s$. The same color denotes the
  same system.}
\label{ftlm-accuracy-f-U}
\end{figure}

It may thus well be that the type of Hilbert space enlargement,
due to growing system size  which leads to the thermodynamic
limit or growing spin quantum number which leads to the
classical limit, is important for the behavior of the estimators
\fmref{E-2-A} and \fmref{E-2-C} at low temperatures.

\section{Discussion and conclusions}
\label{sec-4}

To summarize, we have studied the finite-size scaling of typicality-based trace 
estimators. In these approaches, the trace over the high-dimensional  
Hilbert space is approximated by either (i) a single random state $\ket{r}$ or  
(ii) the average over a set ($R \ll d$) of random vectors. In particular, we 
have focused on the evaluation of thermodynamic observables such as the heat 
capacity and the magnetic susceptibility for various spin models of Heisenberg 
type. Here, the temperature dependence of these quantities has been generated 
by means of a Krylov-space expansion where the random states $\ket{r}$ are used 
as a starting vector for the expansion. 

As a first step, we have studied the full probability distribution of 
expectation values evaluated with respect to single random states. As an 
important result, we have demonstrated that for sufficiently high temperatures 
and large enough system sizes (i.e.\ sufficiently large effective Hilbert-space 
dimension $Z_\text{eff}$), the probability distributions are very well 
described by Gaussians \cite{Reimann2019}. In particular, for 
comparatively high temperatures, our numerical analysis has confirmed that the 
standard deviation of the probability distribution scales as $\delta(O) \propto 
1/\sqrt{Z_\text{eff}}$, and that this width already describes the full 
distribution. 

In contrast, for lower temperatures, we have shown that (i) the probability 
distributions can become non-Gaussian and (ii) the scaling of $\delta(O)$ can 
become more complicated and generally depends on the specific model and 
observable under consideration.  While a larger Hilbert-space dimension often 
leads to an improved accuracy of the random-state approach at low 
temperatures as well, compare the investigation on kagome
lattice antiferromagnets of sizes $N=30$ and $N=42$ in
\cite{SSR:PRB18}, we have also provided examples where this 
expectation can break down for too small $Z_{\mathrm{eff}}$,
compare also \cite{MoT:A19}.

A remarkable example is provided by the spin-$1/2$ 
sawtooth chain with coupling-ratio $|J_2/J_1| = 1/2$. Due to the
(virtually) gapless dense low-energy spectrum at the quantum
critical point, we have found that  
statistical fluctuations remain negligible throughout the entire temperature 
range with only minor dependence on system size (see also 
Ref.~\cite{Rousochatzakis2019} for a similar finding in a spin-liquid model).  
 
In conclusion, we have demonstrated that typicality-based estimators provide a 
convenient numerical tool in order to accurately approximate thermodynamic 
observables for a wide range of temperatures and models. While in some 
cases, even a single pure state is sufficient, the accuracy of the results can 
always be improved by averaging over a set of independently drawn states.

\section*{Acknowledgment}

This work was supported by the Deutsche Forschungsgemeinschaft DFG
(397067869 (STE 2243/3-1); 355031190 
(FOR~2692); 397300368 (SCHN~615/25-1)). 
Computing time at the Leibniz Center in Garching is gratefully
acknowledged. All authors thank Hans De
Raedt, Peter Prelov\ifmmode \check{s}\else \v{s}\fi{}ek, Patrick
Vorndamme, Peter Reimann, Jochen Gemmer as well as Katsuhiro
Morita for valuable comments.


%

\end{document}